\AtBeginDocument{\fontsize{12}{17}\selectfont}

\documentclass{appolb}

\usepackage{graphicx}
\usepackage{amsmath} 
\usepackage{enumerate}

\usepackage{color}
\usepackage[dvipsnames]{xcolor}
\newcommand{\red}[1]{\textcolor{red}{#1}}
\usepackage{datetime2}
\usepackage[english]{babel}
\usepackage{url}
\usepackage{hyperref}
\usepackage{slashed}
\usepackage{caption}
\usepackage{subcaption}
\usepackage{amssymb}

\graphicspath{{plots/}}

\usepackage[mathlines]{lineno}
\begin{document}

\title{\Large{Evidence of isospin-symmetry violation in
high-energy collisions of atomic nuclei:  
\\
Theoretical and phenomenological considerations
}}

\author{
Wojciech Brylinski
\address{Warsaw University of Technology, Warsaw, Poland}
\\[3mm]
Marek Gazdzicki, Francesco Giacosa
\address{Jan Kochanowski University, Kielce, Poland}
\\[3mm]
Mark Gorenstein
\address{Bogolyubov Institute for Theoretical Physics, Kyiv, Ukraine}
\\[3mm]
Roman Poberezhniuk
\address{University of Houston, Houston, Texas, USA}
\\[3mm]
Subhasis Samanta
\address{Department of Physics, School of Applied Sciences, Kalinga Institute of Industrial Technology, Bhubaneswar, Odisha, India}
\\[3mm]
}
\maketitle
\begin{abstract}
Recently, the NA61/SHINE collaboration at the CERN SPS reported evidence of isospin-symmetry violation in high-energy nuclear collisions 
%(Nature Communications \textbf{16}, 2849 (2025)).  
[Nature Commun. \textbf{16}, 2849 (2025)].
The effect was observed in the relative yields of charged and neutral kaons and cannot be explained by known sources of isospin symmetry breaking. In this work, we extend the theoretical and phenomenological aspects of that study.  
We discuss the historical background and introduce the concepts of isospin transformations and symmetry.  
Importantly, we relate isospin symmetry to the QCD flavour symmetry, and we present both conceptual and analytical proofs demonstrating the equality of the mean multiplicities of charged and neutral kaons for an initial ensemble of colliding systems that is invariant under charge-symmetry transformation.
\end{abstract}

%%%%%%%
\section{Introduction}
\label{sec:intro}

\begin{figure}[h!]
\centering

\resizebox{0.9\textwidth}{!}{
  \includegraphics{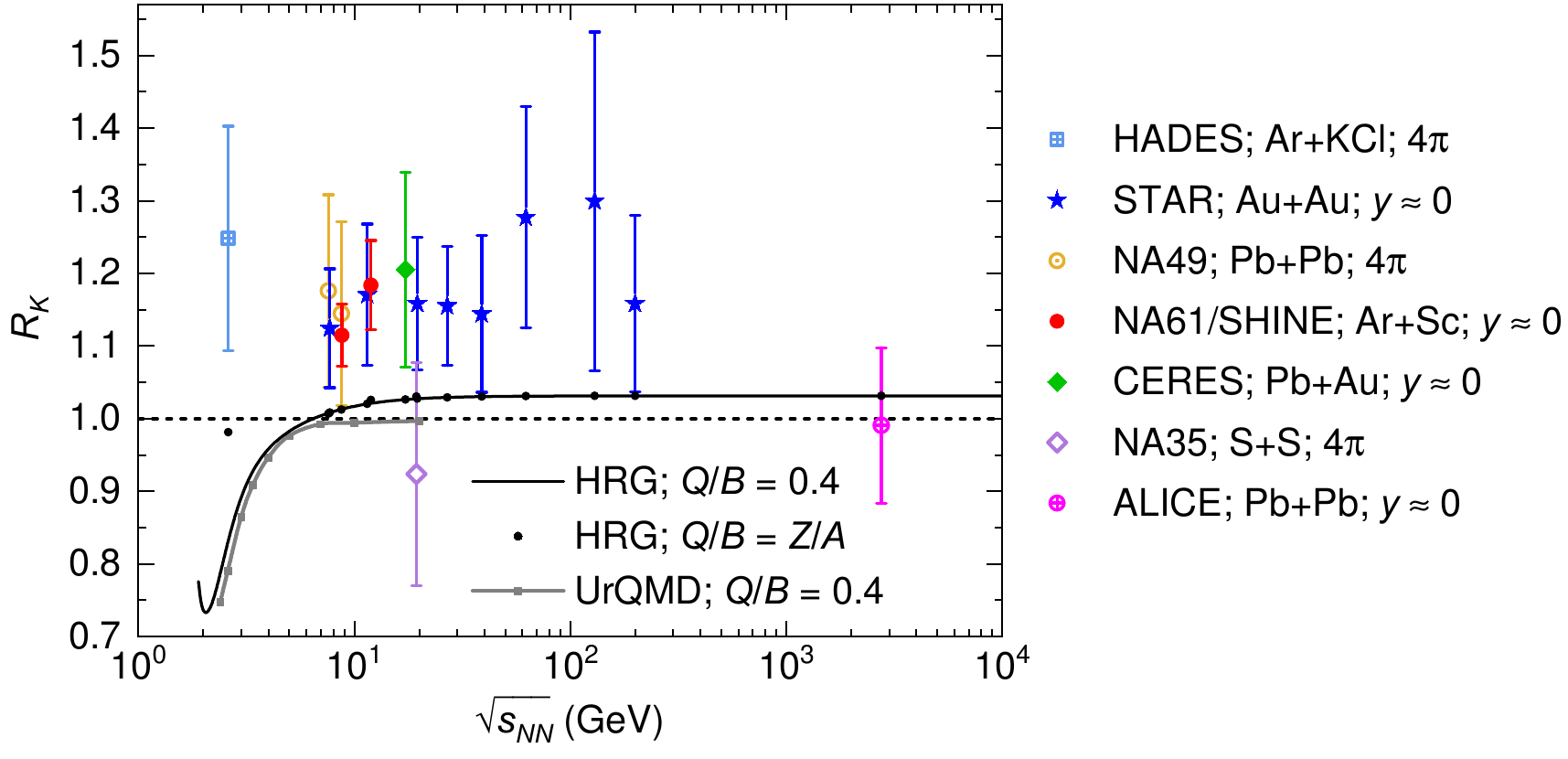}
}
\caption{Ratio $R_K$ of charged to neutral $K$ meson yields in nucleus-nucleus collisions as a function of collision energy. The lines show model predictions; see below for explanations. Figure adapted from Ref.~\cite{NA61SHINE:2023azp} with new preliminary Ar+Sc 8.8~GeV point added~\cite{Balkova:2025dpf}. 
}
\label{fig:NA61_RK}
\end{figure}

In May 2025, the NA61/SHINE collaboration at the CERN SPS published a significant excess of charged over neutral kaon production in high-energy nucleus-nucleus collisions~\cite{NA61SHINE:2023azp}.
For completeness, we recall the key plot of the paper (see Fig.~\ref{fig:NA61_RK}), which shows the ratio of charged to neutral kaons as a function of collision energy. The NA61/SHINE results are presented alongside a compilation of worldwide data.

The largely experimental paper included a theoretical interpretation of the results in terms of evidence for the violation of isospin symmetry beyond known effects. The latter part of the paper was formulated in collaboration with a group of theoretical colleagues.
This paper extends the theoretical and phenomenological aspects of the experimental discovery reported in Nature Communications.

The paper is organised as follows. Section~\ref{sec:history} briefly overviews the history of the isospin concept, whereas an introduction to the isospin transformation and symmetry is presented in Sec.~\ref{sec:isospin}. 
In Sec.~\ref{sec:flavour} we recall that the isospin symmetry follows from the flavour symmetry of up and down quarks. The method used by NA61/SHINE for testing the isospin symmetry by checking the charge-symmetry invariance in 
the ensemble of charge-uniform collisions is elaborated in Secs.~\ref{sec:charge} and~\ref{sec:KOS}. In particular, we present the conceptual and analytical proofs of the method.
In Sec.~\ref{sec:violation}, we estimate the charge-symmetry violation (CSV) within the Hadron Resonance Gas (HRG) model, and we discuss the formally analogous case of $D$ meson production. Finally, in Sec. ~\ref{sec:closing} we present closing remarks and some selected outlook.

\section{Historical background}
\label{sec:history}
In 1932, Heisenberg suggested that the neutron and the proton could be seen as different manifestations of the same particle, the nucleon~\cite{Heisenberg:1932dw}, explaining their similar masses and interaction properties. Wigner introduced the concept of `isotopic spin', or `isospin' of nucleons, following the formal similarity with the more familiar spin~\cite{Wigner:1936dx} and postulated invariance (symmetry) of strong interactions with respect to rotations in the isospin space. 
Today, isospin is an important element of nuclear physics and nuclei classification~\cite{2006NatPh...2..311W}. 
The concept of isospin was promptly extended to the other strongly interacting particles, the hadrons. In particular, Kemmer~\cite{Kemmer:1939zz} applied it to the pion triplet postulated by Yukawa in 1935~\cite{Yukawa:1935xg}. 
In general, when considering the hadron mass distribution, hadrons form groups (isospin multiplets) with members having similar masses but different electric charges. This suggests that strong interactions predominantly determine the masses and strong interaction properties of hadrons, whereas electromagnetic — and, to a much smaller degree, weak — interactions give rise to only small corrections.
For a historical outline, see Refs.~\cite{Kemmer:1982bj, Webb}.

With the continuous discovery of new hadron species, it has become clear that they are not elementary particles of strong interactions.
The quark model of hadron classification proposed by Gell-Mann~\cite{Gell-Mann:1964ewy} and
Zweig~\cite{Zweig:1964jf} in 1964 started a 15-year-long term in which sub-hadronic particles, quarks and gluons were discovered, and a theory of their interactions, 
Quantum Chromodynamics (QCD)~\cite{Fritzsch:1973pi} was established. There are six types (flavours) of quarks: up ($u$), down ($d$), strange ($s$), charm ($c$), bottom ($b$), and top ($t$) with their QCD masses increasing by five orders of magnitude. Each quark carries a colour charge (red, green, or blue), while the gluons are the gauge bosons of QCD and mediate the colour interactions.

Within QCD, hadrons are viewed as confined bags of quarks and gluons. 
%Because of confinement, QCD is based on a local colour gauge symmetry that is hidden in colourless hadrons. 
They are further classified into mesons with integer spin, typically
quark-antiquark states such as pions and kaons, and baryons with semi-integer spin, typically three-quark states such as protons and neutrons. 
However, to date, attempts to derive quantitative QCD predictions for hadron properties and multi-hadron production in high-energy collisions have not been successful. 
Predictions of QCD-based and QCD-inspired models suffer from
large uncertainties. For recent review see Ref.~\cite{Lukashov:2024jej}.
In Sec.~\ref{sec:flavour}, we argue that the exception here is the isospin symmetry and, in particular, its part called charge-symmetry invariance.

Within QCD, the hadron isospin symmetry is considered as a result of the flavour symmetry postulated by QCD. It means that gluons interact `democratically' with all quark flavours. But the results of the interactions depend heavily on the different quark masses. 
However, the two of them, up and down quarks, have masses close to zero ($m_{u}=2.16\pm\ 0.07$ MeV and $m_{d}%
=4.70\pm0.07$ MeV~\cite{ParticleDataGroup:2024cfk})
in comparison to the typical strong interaction scale 
($\Lambda_{QCD} \approx 250$~MeV~\cite{Kneur:2011vi}). Thus, one traces the hadron isospin symmetry to the $ud$-flavour symmetry\footnote{The textbooks refer to this symmetry as the quark isospin symmetry. Here to avoid a possible confusion whether the symmetry is meant at the quark or hadron level, we call it the $ud$-flavour symmetry.} of QCD~\cite{Pal:2014,Donoghue:1992dd}. 
In the limit where the $u$ and $d$ quarks can be considered massless (the so-called chiral limit), the interactions do not depend on their chirality, leading to a larger symmetry called chiral symmetry. The latter is spontaneously broken in the QCD vacuum, generating (quasi-)Goldstone bosons, pions, kaons, and the $\eta$-meson~\cite{Donoghue:1992dd,tHooft:1986ooh}.

Experiments show that the isospin symmetry is well fulfilled in pion-pion and pion-nucleon elastic or quasi-elastic scatterings~\cite {Pennington:2005be,Alarcon:2011zs}. 
%Small deviations in masses and decays within model calculations~\cite{Kovacs:2024cdb,Giacosa:2024epf} and chiral perturbation theory~\cite{Cirigliano:2003gt}.
Moreover, the smallness of isospin-breaking decays (for example $\eta'\rightarrow \pi \pi \pi$~\cite{Gross:1979ur,Serpukhov-Brussels-AnnecyLAPP:1984xib}) is observed.

The above suggests that the isospin symmetry should also be well obeyed in hadron production processes. The results of NA61/SHINE~\cite{NA61SHINE:2023azp} seem to contradict this expectation~\cite{Gazdzicki:1991ih}.

%%%%%%%%%%%%
\section{Isospin transformation and symmetry}
\label{sec:isospin}

In order to formulate the isospin symmetry of strong interactions, one introduces the isospin vector and the isospin transformation. This is done at the level of the system's quantum state, representing either an individual hadron or many hadrons.
The isospin transformation is a rotation of the isospin vector in the isospin space. It is analogous to a rotation of the spin vector in 3-dimensional space, but acting on states related by the (approximate) $SU(2)$ isospin symmetry of the strong interaction—the state multiplets.

More specifically, the isospin operator triplet
$\hat{\boldsymbol{I}}=(\hat{I}_x,\hat{I}_y,\hat{I}_z)$ refers to an $SU(2)$ internal rotation within a given multiplet.
A generic isospin rotation by the angle $\theta_j$ takes place via the operator 
\begin{equation}
\label{eq:it}
\hat{O}=e^{i\theta_{j}\hat{I}_{j}}.
\end{equation}

%\vspace{0.2cm}
Importantly, while isospin $I$ refers to a multiplet of states, $I_z$ is a property of a single state within this multiplet. Moreover, $I_z$ can be inferred from the measured hadron properties using the Nishijima--Nambu--Gell-Mann (NNG) formula~\cite{Nakano:1953zz,Gell-Mann:1956iqa}, which for light hadrons reads:
\begin{equation}
\label{eq:NNG}
I_z = Q - (B + S)/2~,
\end{equation}
where $Q$, $B$, and $S$ are the electric charge, baryon number, and strangeness of the assembly of measured hadrons. In general, a given value of $I_z$ obtained from the experimental data corresponds to a distribution of $I$ values within the range:
\begin{equation}
\label{eq:range}
    I_{\min} \leq I \leq I_{\max}~,
\end{equation}
where $I_{\min} = |I_z|$ and $I_{\max}$ is the maximum total isospin when all isospin vectors are aligned. 
This makes the experimental test of full isospin symmetry significantly more difficult than the tests of its $I_z$-related sub-symmetry, the so-called charge symmetry. \\
The charge-symmetry transformation is obtained by setting \mbox{$\theta_1 = \theta_3=0$} and $\theta_2=\pi$ in Eq. \ref{eq:it}. 
The corresponding operator reads,
\begin{equation}
\label{eq:cst}
\hat{C}=e^{i\pi \hat{I}_{y}}.  
\end{equation}
When applied to a hadronic state, this operator inverts the $I_z$ component of isospin.

\vspace{0.2cm}
Let us start by considering examples for a single hadron.
Here, the measurement of hadron mass identifies the multiplet, allowing us to assign both $I$ and $I_z$ to the underlying state of each measured hadron.\\
The simplest isospin multiplet is realised for isospin singlets, also denoted as isoscalar. For example, there is only one isospin state for the meson $\omega$ and for the $\Lambda$ hyperon ($I = 0$).\\
The pair proton ($p$) and neutron ($n$) forms an `isodoublet' ($I = 1/2$), where  $p$ has eigenvalue $I_{z}=1/2$ and $n$ has $I_{z}=-1/2$. This also holds for the pair $K^+$ and $K^0$. 
Thus, an isospin transformation on the doublet $(p,n)^T$ mixes proton and neutron and is obtained by setting $\hat{I}_{j} = \tau_j/2$, with $\tau_j$ being the Pauli matrices.
The antinucleon isodoublet $(-\bar{p},\bar{n})^T$ transforms just as the nucleon one (note the minus sign~\cite{Halzen:1984mc}). \\
An isospin triplet ($I=1$), for example, the pion triplet, is represented by a column vector:
\[
\vec{\pi} =
\begin{pmatrix}
-\pi^+ \\[6pt]
\pi^0 \\[6pt]
\pi^-
\end{pmatrix}~.
\]
Then an isospin transformation is described by identifying the isospin operators $\hat{I}_j$ with the $3 \times 3$ matrices $\hat{J}_j$.  
These matrices realise the spin-1 irreducible representation of $SU(2)$, also called the adjoint representation.\\
The charge-symmetry transformation, Eq.~\ref{eq:cst}, replaces a hadronic state $|I, I_z\rangle$ by its multiplet partner $|I, -I_z\rangle$. This results in swapping between measured hadrons as $p \leftrightarrow n$, $\pi^+ \leftrightarrow \pi^-$, $K^+ \leftrightarrow K^0$, $\bar{K}^0 \leftrightarrow K^-$, etc.

We turn now to assemblies of two or more hadrons and atomic nuclei.
The selected examples will be used later in the paper.
\begin{enumerate}[(A)]
\item Firstly, we consider an example of two nucleons.  
For two protons, using Eqs.~\ref{eq:NNG} and~\ref{eq:range}, one gets 
$I_z = +1$ and $I = 1$ ($I_{\min} = I_{\max} = 1$).  
Similarly, for the two--neutron state, one gets $I_z = -1$ and $I = 1$.  
Thus, in both cases $I$ is uniquely defined.  
However, for a proton-neutron state, the situation is more complicated: 
here $I_z = 0$, but $I_{\min} = 0$ and $I_{\max} = 1$, allowing for two possible 
values of the total isospin, $I=0$ and $I=1$.  
Consequently, the $|p,n\rangle$ and $|n,p\rangle$ states are not eigenstates of 
total isospin, as they are linear combinations of the singlet ($I=0$) and triplet ($I=1$) components.  
The charge--symmetry transformation changes the $|p,p\rangle$ state into the 
$|n,n\rangle$ one and vice versa, and similarly maps $|p,n\rangle$ into $|n,p\rangle$.  
This means that an ensemble of $|p,p\rangle$ and $|n,n\rangle$ states with an equal 
number of each is invariant under charge symmetry.

\item
Secondly, we discuss the case of an atomic nucleus with equal numbers of protons and neutrons\footnote{In terms of conserved quantities, the condition $Z = N$ is given by $Q/B=1/2$, where $Q=Z$ and $B=A$ are the nuclei's electric charge and baryon number, respectively.}, $Z = N$, $ A = Z + N$. For a single nucleus one gets $I_z = 0$ and $0 \leq I \leq (Z + N)/2$. Thus, the state of two nuclei $|A, A \rangle$ has $I_z = 0$. It is transformed into itself by the charge-symmetry transformation independently of its $I$ value in the phenomenology of nuclear physics; these states are also charge-symmetry invariant. 
This property will later be used to test charge-symmetry invariance in nucleus-nucleus collisions. 
From phenomenology of nuclear physics follows that, for almost all nuclei of the $Z = N$ type, the total isospin is zero~\cite{Lenzi2009}. Then the $|A, A \rangle$ state has $I_z = I = 0$.
This would allow the extension of tests of isospin symmetry beyond charge symmetry by studying nucleus-nucleus collisions.
\item
Thirdly, we consider  $|\pi, A \rangle$ states for $Z = N$ nuclei. The $|\pi^+, A \rangle$ state has $I_z = 1$, whereas the $|\pi^-, A \rangle$ state has $I_z = -1$. 
Similar to the $|p, p \rangle$ and $|n, n \rangle$ states, the charge-symmetry transformation swaps $|\pi^+, A \rangle$ and $|\pi^-, A \rangle$ states.
The ensemble of equal numbers of $|\pi^+, A \rangle$ and $|\pi^-, A \rangle$ states is invariant under charge-symmetry transformation, whereas each state is not. 
\end{enumerate}

\vspace{0.2cm}
Isospin symmetry refers to the invariance of the strong interaction between hadrons under global $SU(2)$ rotation in isospin space. 
The transition probability from a given initial state to a given final state equals the transition probability between the states obtained by applying the isospin transformation to these states.
This symmetry has many empirical consequences, for example:
\begin{enumerate}[(i)]
    \item The elastic cross sections of $p+p$ and $n+n$ scattering should be the same.
    \item The mean multiplicity of $K^+$ mesons produced in $p+p$ interactions is equal to the mean multiplicity of $K^0$ mesons produced in $n+n$ interactions,
    \begin{equation}
    \label{eq:kaons_NN}
    \langle K^+ \rangle_{pp} = \langle K^0 \rangle_{nn}~.
    \end{equation}
\end{enumerate}

Many additional relations of this kind follow from the same symmetry principle.

%%%%%%%%
\section{Isospin symmetry from \textit{ud}-flavour symmetry}
\label{sec:flavour}
%%%%%%%%
%\subsection*{Isospin symmetry from quark flavour symmetry}

In QCD, the fundamental degrees of freedom are quarks and gluons. 
Restricting ourselves to the two lightest flavours ($u$ and $d$), the QCD Lagrangian reads
\begin{equation}
\mathcal{L}_{\text{QCD}} = 
\bar u \left(i\slashed{D} - m_u\right) u \;+\;
\bar d \left(i\slashed{D} - m_d\right) d \;+\; \ldots
\text{ ,}
\end{equation}
where $\slashed{D}=\slashed{\partial}-ig_{QCD}\gamma^{\mu} A_{\mu}$, with $A_{\mu}$ being the gluon field. 
If the up and down quarks were mass-degenerate ($m_u = m_d$), this part of the Lagrangian 
would be invariant under global $SU(2)$ flavour rotations mixing the $u$ and $d$ states,
\begin{equation}
\begin{pmatrix} u \\ d \end{pmatrix} \;\longrightarrow\;
\hat{O} \begin{pmatrix} u \\ d \end{pmatrix}, 
\qquad \hat{O} \in SU(2) \text{.}
\end{equation}
This $SU(2)$ $ud$-flavour (quark isospin) symmetry is only approximate because $m_u \neq m_d$.
The pair $u$ and $d$ forms an `isodoublet', where the quark $u$
has eigenvalue $I_{z}=1/2$ and $d$ has $I_{z}=-1/2$ (see e.g. Ref.~\cite{Pal:2014} and the quark summary in the PDG~\cite{ParticleDataGroup:2024cfk}). 
Thus, the $ud$-flavour transformation on the doublet $(u,d)^T$ mixes both quarks and is obtained by setting $\hat{I}_{j} = \tau_j/2$, with $\tau_j$ being the Pauli matrices.
The antiquark isodoublet $(-\bar{d},\bar{u})^T$ transforms just as the quark one (note the minus sign~\cite{Halzen:1984mc}). 
Since hadrons are colour-singlet bound states of quarks and gluons, they inherit this symmetry: the spectrum of hadrons is arranged into isospin multiplets such as the nucleons ($p,n$) or the pions ($\pi^+,\pi^0,\pi^-$). 

%The symmetry is not exact, being broken by the 
%quark mass difference $m_u \neq m_d$ and by electromagnetic interactions that %distinguish $u$ and $d$ quarks through their different electric charges. These %effects account for the small but observable mass splitting within hadronic %isospin multiplets, for example, $m_p - m_n$.

\vspace{0.2cm}
The fact that the $ud$-flavour transformation translates into the one for isodoublets can be seen by using, for example, microscopic currents used in 
lattice calculations~\cite{Dudek:2013yja}, Bethe-Salpeter approaches~\cite{Maris:2003vk} and effective models of QCD~\cite{Hatsuda:1994pi,Giacosa:2024epf}.

The $ud$-flavour symmetry of the QCD Lagrangian means that certain Noether currents and charges are conserved. Namely, for the global $SU(2)$ transformation of the light quark doublet, the Noether currents are given by
\[
J_{i}^{\mu} = \bar q\,\gamma^\mu \tfrac{\tau_i}{2}\, q  \text{ ,}
\]
where $q^T=(u,d)$. Their divergence follows from the QCD equations of motion,
\[
\partial_\mu J_i^{\mu}
= \bar{q}\, i \!\stackrel{\leftrightarrow}{\slashed{D}} \,\frac{\tau_i}{2}\, q
= (m_u - m_d)\,\bar{q}\,\frac{\tau_i}{2}\, q \, .
\]
Hence, in the exact $ud$-flavour symmetry limit $m_u = m_d$, the divergences vanish and the currents are conserved. Turning back to hadrons, isospin symmetry implies that only transitions preserving the total isospin $I$ and its third component $I_z$ are allowed. This property is the already mentioned isospin conservation.

In particular, when the initial and final states are eigenstates of both $I$ and $I_z$, the consequences of the symmetry are straightforward. However, in heavy-ion collisions, the initial state is typically an eigenstate of $I_z$ (since $Q$, $B$, and $S$ are conserved event-by-event) but not necessarily an eigenstate of $I$. In such cases, isospin conservation must be interpreted with some care.

If the initial state is a \emph{classical mixture} of components with different total isospin values $I$, the symmetry applies separately to each component, and the composition of the mixture remains unchanged. If, instead, the initial state is a \emph{quantum superposition} of several states with distinct values of $I$, the interaction cannot mix them. Consequently, the relative weights of the different $I$ components of the superposition remain fixed in time, and observables depending on $I$ are conserved at the level of expectation values.

\iffalse
Hence, in the exact $ud$-flavour symmetry limit $m_u=m_d$, the divergences vanish, and the currents are conserved.
Turning back to hadrons, isospin symmetry implies that only transitions preserving total isospin $I$ and its $z$-component $I_z$ are allowed. This property is known as isospin conservation. 

FG: I will argue like that. For $I$ and $I_z$ fixed  (i.e. eigenstates), the situation is pretty clear. Yet, in HIC, the initial state is an eigenstate of $I_z$ but not necessarily of $I$. In this case, conservation means:
(i) classical superposition....trivial
(ii) quantum superposition...then $\langle I \rangle$ is conserved.

\red{explain what is meant here: $I_z$ is conserved event-by-event (Q, B, S are conserved), but what about $I$? What about states being superpositions of different $I$ values, for example $np$?} is well supported by \red{empirical data~\ref{} (add reference)}. 
\fi

The QCD $u$ and $d$ quark masses extracted from the experimental data\footnote{
The QCD `current' quark masses $m_u= 2.16 \pm 0.07$ MeV and $m_d = 4.70 \pm 0.07$ are not measured directly but are obtained by fitting experimental data on hadron properties \cite{ParticleDataGroup:2024cfk}. Light-quark masses are fitted within lattice QCD and chiral perturbation theory, while heavy-quark masses come from perturbative QCD analyses of quarkonium and collider data. They are always quoted in a particular renormalisation scheme.} are different but small.
They are much smaller than $\Lambda_{QCD}$. Thus, one argues that $ud$-flavour symmetry is well obeyed. 
Yet, a closer inspection of the link between the approximate $ud$-flavour and isospin symmetries shows that other features of QCD are also needed to obey isospin symmetry approximately.
We discuss them below.

\begin{enumerate}[(a)]
\item 
Neglecting at first the Goldstone bosons of QCD (pions, kaons,...), the masses of the light
hadrons can be successfully described by the constituent quark model approach with $m_{u}^{\ast}\simeq m_{d}^{\ast}\sim
\Lambda_{QCD}\sim250 \text{ MeV}$, see Refs.~\cite{Godfrey:1985xj,Lucha:1991vn}. For instance, the masses of conventional $\bar{q}q$ mesons, such as the $\rho(770)$ meson, can be described as bound states of a constituent quark and a constituent antiquark: their mass is proportional to the sum of the constituent quark masses
that build it. 
As shown by microscopic models, such as the Nambu Jona-Lasinio one~\cite{Hatsuda:1994pi} and Bethe-Salpeter approaches~\cite{Fischer:2006ub}, the difference of the constituent masses is proportional to the difference of QCD
masses, $m_{d}^{\ast}-m_{u}^{\ast}\simeq m_{d}-m_{u}$,
thus suppressed. For example, the physical $\rho(770)$-meson masses lead to the mass difference $(m_{\rho^+}-m_{\rho^0})=-0.7 \pm 0.8$~MeV.
Within this context, one can easily
understand that $m_{u,d}\ll\Lambda_{QCD}$ implies that isospin symmetry is
well fulfilled. The symmetry is also well respected in
decays and scattering properties; see~Refs.~\cite{Giacosa:2024epf,Kovacs:2024cdb}.
\item
The pseudoscalar Goldstone bosons (pions, kaons, and $\eta(547)$) are special, since their mass squared is proportional to the QCD quark masses that form them. 
For charged pions one has $m_{\pi^{\pm}}^{2}\propto(m_{u}+m_{d}).$   For the four
kaons, central to this work, one has:
\begin{equation}
m_{K^{-}}^{2}=m_{K^{+}}^{2} = c_K(m_{u}+m_{s})\text{, }m_{\bar{K}^{0}}%
^{2}=m_{K^{0}}^{2}= c_K(m_{d}+m_{s})\text{ ,}%
\end{equation}
where $c_K$ is an appropriate constant that depends on the quark condensates, e.g. \cite{Giacosa:2024epf}.
Thus, the similarity of $K^{+}$ and $K^{0}$ masses (and similarly $K^{-}$ and $\bar{K}%
^{0})$ is reflected in the QCD masses of up, down, and strange quarks. In particular, the strange quark mass has to be large enough, $m_{s}%
=93.5\pm0.8$~MeV~\cite{ParticleDataGroup:2024cfk}.
For a cross-check, one may evaluate the ratio 
\begin{equation}
\frac{m_{K^{+}}}{m_{K^{0}}}=\sqrt{\frac{m_{u}+m_{s}}{m_{d}+m_{s}}}%
=0.9870\pm0.0005 
\text{ ,}
\end{equation}
whose central value is close to the corresponding central experimental value $0.99209\pm
0.00004$.
Yet, due to the very small errors, these values are not compatible. The reason is the small but decisive role of the electromagnetic interaction \cite{Donoghue:1996zn}, which slightly increases the mass of $K^+$ (both constituents $u$ and $\bar{s}$ have positive charge), but slightly decreases the mass of $K^0$ (the constituents $d$ and $\bar{s}$ have opposite charge).

Using the statistical approach, one can estimate the ratio of mean multiplicities of charged and neutral kaons caused by the mass difference as
\begin{equation}
\frac{\langle K^{+} \rangle}{\langle K^{0} \rangle} \approx e^{-\frac{m_{K^+} - m_{K^0}{}}T } \approx 1.03~,
\end{equation}
where $T = 150$~MeV is the temperature parameter of the kaon production process. This estimate assumes direct kaon production. The detailed estimate of the effect, including resonance decays, is presented in Sec.~\ref {sec:violation}. It is about several $\%$ at $\sqrt{s}\simeq 10-20$ GeV. Note that this estimate includes both strong and electromagnetic effects that modify vacuum masses of hadrons and their branching ratios (assuming they have been measured).

\item
A special remark about the neutral members of the pseudoscalar
nonet is needed. The neutral pion carries a mass $m_{\pi^{0}}^{2}\propto(m_{u}+m_{d})$
just as the charged ones. The isoscalar members $\eta_{N}=(u\bar
{u}+d\bar{d})/\sqrt{2}$ and $\eta_{s}=$ $s\bar{s}$ develop a mass $m_{\eta_{N}}%
^{2}\propto(m_{u}+m_{d})+2c_{A}$ and $m_{\eta_{S}}^{2}\propto m_{s}+c_{A}$, as
well as a mixing proportional to $\sqrt{2}c_{A}$. Here $c_{A}$ is a
constant proportional to the $U(1)_A$ chiral (or axial) anomaly.
The latter refers to a specific $U(1)_A$-chiral-phase symmetry of quarks that is a symmetry of the classical QCD, which is broken by gluonic quantum 
fluctuations~\cite{tHooft:1986ooh,tHooft:1976rip}.
Since the coefficient $c_{A}$ is large (in fact dominant compared to the contributions of the QCD quark masses), 
the masses of the isoscalar mesons $\eta_{N}$ and $\eta_{S}$ become much larger than that of the neutral pion $\pi^0$. The corresponding states effectively decouple from it. Instead, $\eta_{N}$ and $\eta_{S}$ mix to form the physical 
mesons $\eta(547)$ and $\eta^{\prime}(958)$. The latter is particularly massive and cannot be regarded as a Goldstone boson.
The neutral pion $\pi^0$ is almost degenerate with the charged states $\pi^{\pm}$,
forming an isospin triplet. As noted in Refs.~\cite{Gross:1979ur,Pisarski:1983ms}, the situation would have been different if the chiral anomaly had been small. In this case, the $\pi^{0}$ and the $\eta_{N}$ states would be almost mass degenerate. A residual $\pi^{0}$-$\eta_{N}$ small mixing term, proportional to $(m_{d}-m_{u})$, would generate a large mass splitting. The lighter state would have a mass
of about $77$~MeV. Thus, pions would not form an isospin triplet of almost degenerate masses.
Colloquially, we may say that the large chiral anomaly protects isospin symmetry
at the level of the pions\footnote{This is not the case for other nonets. Even if the chiral anomaly is non-zero, it is never the dominating component as for the $\eta^{\prime}$-meson~\cite{Giacosa:2023fdz,Giacosa:2017pos}.}. Importantly, the charge symmetry is not affected by the chiral anomaly.
\end{enumerate}

\bigskip
The relation between the phenomenological isospin and the QCD $ud$-flavour symmetries can be summarised as follows:
\begin{enumerate}[(A)]
\item 
The exact flavour symmetry of up and down quarks ($m_u = m_d$) results in the exact isospin symmetry of hadrons.
\item 
Taking the QCD quark masses leads to a small breaking of the $ud$-flavour symmetry, which in turn causes a small breaking of isospin symmetry. In particular, the charge symmetry is weakly affected.
\item 
The magnitude of isospin symmetry violation cannot be calculated exactly in QCD.
For mean kaon multiplicities, the estimate is approximately several per cent; see Sec.~\ref{sec:violation} for details.
\end{enumerate}

%%%%%%%%%%
\section{Charge-uniform ensembles and kaons}
\label{sec:charge}

The simplest conceptual test of charge symmetry in kaon production is given by Eq.~\ref{eq:kaons_NN}, which for completeness we repeat here:
$\langle K^+ \rangle_{pp} = \langle K^0 \rangle_{nn}~$.
One needs to measure the mean multiplicity of $K^+$ mesons in proton-proton interactions and compare it with the corresponding measurement for $K^0$ mesons in neutron-neutron interactions. The $K^+$ ($u\bar{s}$) and $K^0$ ($d\bar{s}$) mesons are charge symmetry partners. At the quark level, they transform by swapping up and down quarks. Importantly, kaons are copiously produced in high-energy collisions. \\
However, this test is almost impossible to perform experimentally.
Neither neutron beams nor neutron targets are easy to produce. Measurements of $K^0$ production are possible only through its strong interactions with the detector material. It is also a challenging task. In view of these difficulties and based on previous consideration~\cite{Gazdzicki:1991ih}, NA61/SHINE performed the test of 
the charge-symmetry invariance in the kaonic sector, measuring the ratio
of charged ($K^+ + K^-$) to neutral ($K^0 + \bar{K}^0 = 2 \cdot K^0_S$) kaons in nucleus-nucleus ($Z \approx N)$ collisions. In this and the following sections, we elaborate on this test method.

%%%%%%%%
\subsection{Conceptual proof}
\label{sec:uni_concept}

The conceptual proof of the correctness of the NA61/SHINE test reads:
\begin{enumerate}[(i)]
\item 
Assume the ensemble of initial states is invariant under charge-symmetry transformation.
It means the ensemble has an equal population of charge-symmetry partners.
It is the charge-uniform ensemble.
Here, the important example is the ensemble of two colliding nuclei, each with equal numbers of protons $Z$ and neutrons $N$, $Z = N = A/2$. For instance, this can be an ensemble of $^{16}O$ $+$ $^{16}O$ states. 
As discussed in Sec.~\ref{sec:isospin}, point (B), each state of the ensemble has $I_z = 0$, and thus it is invariant under charge transformation. Obviously, the ensemble of these states is charge-uniform.
The second example is the ensemble of an equal number of $\pi^+ + C$ and $\pi^- + C$ systems. As discussed in Sec.~\ref{sec:isospin} point~(C), the states have $I_z = \pm 1$. Thus, the overall ensemble is charge-uniform, despite each initial state, taken separately, not being charge-symmetry invariant.
The experimental study of $^{16}O$ $+$ $^{16}O$ 
as well as $\pi^+ + C$ and $\pi^- + C$ collisions was recently initiated by 
NA61/SHINE~\cite{Grebieskow:2941113}.
\item 
Assume that strong interactions are charge-symmetry invariant; see Sec.~\ref{sec:isospin} for details.
\item 
From the assumptions (i) and (ii), it trivially follows that the ensemble of final states is charge-symmetry invariant (charge uniform). A charge-symmetric initial ensemble yields a charge-symmetric final ensemble. 
This means that the mean multiplicities of charge-symmetry partners have to be equal,
\begin{equation}
\label{eq:kplus}
\langle K^{+} \rangle = \langle K^{0} \rangle \text{ , }%
\end{equation}
and 
\begin{equation}
\label{eq:kminus}
\langle K^{-} \rangle = \langle \bar{K}^{0} \rangle \text{ . }%
\end{equation} 
Thus,
\begin{equation}
\label{eq:chn}
\langle K^{+} \rangle + \langle K^{-} \rangle = 
\langle K^{0} \rangle + \langle \bar{K}^{0} \rangle~,
\end{equation}
where, as shown in Sec.~\ref{sec:KOS}, the sum of neutral kaon multiplicities can be replaced by the $2\langle  K^0_S \rangle$.
\end{enumerate}

%%%%%%%%
\subsection{Analytical proof}
\label{sec:uni_operator}

Here, we present an alternative proof of Eqs.~\ref{eq:kplus} and~\ref{eq:kminus} based on the quantum-mechanical formalism. For simplicity, we restrict to the A+A collisions with $N=Z=A/2$.  

\iffalse
\\\red{Within the framework, consider two cases:\\- the A+A case:\\\vspace{0.2cm}$\hat{C} \left\vert \Psi_{n}\right\rangle\left\langle \Psi_{n}\right\vert  \hat{C}^{\dagger} = \left\vert \Psi_{n}\right\rangle\left\langle \Psi_{n}\right\vert $ and \\ 
\vspace{0.2cm}
- the $\pi\pm$ + A case\\
\vspace{0.2cm}
$\hat{C} \left\vert \Psi_{n}\right\rangle
\left\langle \Psi_{n}\right\vert  \hat{C}^{\dagger} = 
\left\vert \Psi^*_{n}\right\rangle
\left\langle \Psi^*_{n}\right\vert $ with $p_n$ and\\ 
\vspace{0.2cm}
$\hat{C} \left\vert \Psi^*_{n}\right\rangle
\left\langle \Psi^*_{n}\right\vert  \hat{C}^{\dagger} = 
\left\vert \Psi_{n}\right\rangle
\left\langle \Psi_{n}\right\vert $ with $p^*_n = p_n$~,\\
\vspace{0.2cm}
where $^*$ labels the charge-symmetry partner.
}
\fi

In terms of the statistical
operator $\hat{\rho}=\sum_{n}p_{n}\left\vert \Psi_{n}\right\rangle
\left\langle \Psi_{n}\right\vert $ describing the initial state of A+A
collisions of two nuclei with $N=Z=A/2$, and thus $I_{z}=0$, invariance under charge-transformation $\hat{C}$ implies the equality
$\hat{C}\hat{\rho}\hat{C}^{\dagger}=\hat{\rho}$. Here, $p_{n}$
represents the probability for a specific initial state $\left\vert \Psi
_{n}\right\rangle $. 
As usual, $Tr\left[  \hat{\rho}\right]  =\sum_{n}p_{n}=1$.

The time-evolution of the statistical operator reads $\hat{\rho}(t) = \hat{U}(t)\hat{\rho}\hat{U}(t)^{\dagger}$, with $U(t)=e^{-i\hat{H}_{QCD} t}$. 
We introduce the kaonic number operators $\hat{N}_{K^{+}}$ and $\hat{N}_{K^{0}}$, which are related by charge- symmetry operator as $\hat
{N}_{K^{+}} = \hat{C}^{\dagger}\hat
{N}_{K^{0}}\hat{C}$. Finally using the charge symmetry of strong interactions, $[U(t),\hat{C}] =0$, the kaon mean multiplicities at a given time $t \geq 0$ read:
\begin{align}
\langle K^{+}\rangle &  =  Tr\left[\hat{\rho}(t)\hat{N}_{K^{+}%
}\right] =Tr\left[U(t)  \hat{\rho}U^{\dagger}(t)\hat{N}_{K^{+}%
}\right]  =Tr\left[ U(t) \hat{\rho}U^{\dagger}(t)\hat{C}^{\dagger}\hat
{N}_{K^{0}}\hat{C}\right] \nonumber\\
&  = Tr\left[ U(t) \hat{C}\hat{\rho}\hat{C}^{\dagger}U^{\dagger}(t)\hat{N}_{K^{0}%
}\right]  =Tr\left[ U(t) \hat{\rho}U^{\dagger}(t)\hat{N}_{K^{0}} \right] \nonumber\\ 
&  = Tr\left[\hat{\rho}(t)\hat{N}_{K^{0}%
}\right]
=\langle K^{0}\rangle\text{ ,}\label{eq:stat}%
\end{align}
obeying Eq.~(\ref{eq:kplus}) for the ensemble of A+A collisions at any later time $t$. Similarly, one proves Eq.~\ref{eq:kminus}, $\langle K^{-}\rangle=\langle\bar{K}^{0}\rangle$.

%%%%%%%%
\subsection{Comments on isospin-uniform ensembles}
\label{sec:uni_comm}

The conceptual proof (Sec.~\ref{sec:uni_concept}) of the relations (\ref{eq:kplus}) and (\ref{eq:kminus}) shows similarity to the setup considered by Shmushkevich~\cite{Shmushkevich:55,Shmushkevich:56} in the mid 1950s.
The setup allowed him to formulate an easy method for relating cross-sections for various processes, following from the invariance of strong interactions under isospin transformations; see also Refs.~\cite{MacFarlane:1965wp,1982AmJPh..50..748W,Pal:2014}. 
The main idea is that an initial `isospin-uniform' ensemble of hadronic states (that is, one with an equal mean number of each member of any isospin multiplet)
evolves into an isospin-uniform final-state ensemble. 

Here we illustrate the Shmushkevich method considering a simple example of $\rho(770)$ ($I = 1$) meson decays.
If initially we have equal number of $\rho(770)^+$, $\rho(770)^0$ and $\rho(770)^-$ independent states, the corresponding statistical operator reads:
\begin{equation}
\hat{\rho}=\frac{1}{3}(\left\vert \rho(770)^{+}\right\rangle \left\langle \rho(770)
^{+}\right\vert +\left\vert \rho(770)^{0}\right\rangle \left\langle
\rho(770)^{0}\right\vert +\left\vert \rho(770)^{-}\right\rangle \left\langle
\rho(770)^{-}\right\vert) \text{ . }%
\end{equation}
As such, it is invariant under any isospin transformation $\hat{O}=e^{i\theta
_{j}\hat{I}_{j}}$ with $\hat{O}\hat{\rho}\hat{O}^{\dagger}=\hat{\rho}$ for an arbitrary
choice of the angles $\theta_j.$ With this initial ensemble, the multiplicities of the members of any isospin multiplet must be equal at all times $t$. 
%For an ensemble of $N$ independent $\rho$ mesons, the uniform state is simply the tensor product of single--meson ensembles, \begin{equation} \hat{\rho}^{(N)} = \hat{\rho}\otimes \hat{\rho}\otimes \cdots \otimes \hat{\rho} \quad (N \text{ times})~. \label{sqrtc} \end{equation} Next, let us take into account an initial isospin-uniform ensemble, such as the one of $\rho$-mesons described above. We may easily show that the final state is also uniform. As a concrete example, let us consider pion multiplicities. 
This point can be shown by considering the isospin transformation $\hat{C}^{1/2} = e^{iI_y \pi/2}$,
leading to\footnote{Under this transformation, $u\rightarrow (u+d)/\sqrt{2}$, $d\rightarrow (-u+d)/\sqrt{2}$. Note, a number operator for a certain isospin multiplet member $r$ is given by $\hat{N}_r= \sum_{\mathbf{p}}a(\mathbf{p},r)^{\dagger}a(\mathbf{p},r)$, with $\mathbf{p}$ being the momentum. Hence, a transformation with arbitrary $\theta_3$ but with  $\theta_{1,2}=0$ assures that all expectations of mixed products $\langle a(\mathbf{p},r)^{\dagger}a(\mathbf{p},l)\rangle$ vanish for $r \neq l$. Then, under $\hat{C}^{1/2}$, the operator $N_{\pi^+}$ leads to Eq. \ref{sqrtc}.}   
\begin{align}
\langle\pi^{+}\rangle  = \frac{1}{4}\langle\pi^{+}\rangle + \frac{1}{2}\langle\pi^{0}\rangle + \frac{1}{4}\langle\pi^{-}\rangle  \text{ .} \label{sqrtc}
\end{align}
Utilising charge-symmetry, $\langle\pi^{+}\rangle  = \langle\pi^{-}\rangle$, one gets $\langle\pi^{+}\rangle  = \langle\pi^{0}\rangle$.

The discussion above is general. An initial ensemble $\hat{\rho}$ being
isospin-uniform at $t=0$, 
\begin{equation}
\hat{\rho}\text{ uniform }\longleftrightarrow\text{ }\hat{O}\hat{\rho}\hat
{O}^{\dagger}=\hat{\rho}\text{ }\forall\text{ }\hat{O}
\text{ ,}
\end{equation}
stays isospin-uniform at any later time:
\begin{equation}
\hat{\rho}(t)=U(t)\hat{\rho}U^{\dagger}(t)\text{ is such that }\hat{O}%
\hat{\rho}(t)\hat{O}^{\dagger}=\hat{\rho}(t)
\text{ }
\forall t \geq 0
\text{ ,}
\end{equation}
because $[  U(t)=e^{-iH_{QCD}t},\hat{O}]  = 0$ is a consequence of
isospin symmetry of the underlying Hamiltonian.

In a nucleus-nucleus collision, the initial states are, in general, not isospin-uniform, see Sec.~\ref{sec:isospin}.
Only if both nuclei have vanishing isospin, the initial state has $I=0$ and is trivially uniform\footnote{Within our vector meson analogy, the case with $I=0$ corresponds to $\hat{\rho} =  \left\vert \omega(782)\right\rangle \left\langle \omega(782)\right\vert $, where $\left\vert \omega(782)\right\rangle$ is an isosinglet state \cite{ParticleDataGroup:2024cfk}. }. 
For example, the ensemble of $I=1$ and $I_z=0$ states is not isospin-uniform. Namely, the uniformity would require having, besides the $I_z =0$ member, also the ones with $I_z = \pm1$, which, however, are not present in the system since only $I_z = 0$ is allowed for $N=Z$ nuclei. Thus, we cannot predict that each member of an isospin multiplet is produced with equal probability (for example, one cannot show that $\langle \pi^{+}\rangle=\langle \pi^{0}\rangle$).

Using the $\rho(770)$-meson example, the case encountered in nucleus-nucleus collisions with $Q/B=1/2$ and $I=1$ is analogous to the charge-invariant statistical operator of the initial ensemble, the $\hat{\rho}=\left\vert \rho^{0}(770)\right\rangle \left\langle
\rho^{0}(770)\right\vert$, where only the member with $I_z = 0$ is retained. This implies that the final ensemble of states is also
charge-symmetry invariant, hence $\langle\pi^{+}\rangle=\langle\pi^{-}%
\rangle,$  $\langle K^{+}\rangle=\langle K^{0}\rangle,$ $\langle K^{-}%
\rangle=\langle\bar{K}^{0}\rangle,$ $\langle p\rangle=\langle n\rangle$, but it does not imply that $\langle\pi^{+}\rangle=\langle\pi^{0}\rangle$. (In fact, $\rho^0(770)$ decays solely into $\pi^+ \pi^-$, implying no $\pi^0$ production.)

In summary, the initial ensemble of nucleus-nucleus collisions with an equal number of protons and neutrons is the charge-uniform ensemble of states. It thus leads to the charge-uniform ensemble of final states. In general, these ensembles do not fulfil the requirement of isospin-uniform ensembles that is necessary for the Shmushkevich method to be valid, not even for $Q/B=1/2$. Nevertheless, charge-symmetry invariance of the initial charge-uniform ensemble is sufficient for having the same mean multiplicities of charged and neutral kaons. 

\section{Neutral kaons and $R_K$}
\label{sec:KOS}

The neutral kaons are eigenstates of the strong interaction and oscillate ($K^0 \leftrightarrow \bar{K}^0$) due to the weak interaction. The weak eigenstates are the long-lived `K-long' $K_{L}^{0}$ and short-lived `K-short' $K_{S}^{0}$, which decay into three and two pions with mean lifetimes of $15~m/c$ and 2.8~cm/$c$, respectively. By neglecting at first the small CP violation (see below), the weak and strong kaonic eigenstates are related by:
\begin{equation}
\left(
\begin{array}
[c]{c}%
\left\vert K_{S}^{0}\right\rangle \\
\left\vert K_{L}^{0}\right\rangle
\end{array}
\right)  =\frac{1}{\sqrt{2}}\left(
\begin{array}
[c]{cc}%
1 & 1\\
1 & -1
\end{array}
\right)  \left(
\begin{array}
[c]{c}%
\left\vert K^{0}\right\rangle \\
\left\vert \bar{K}^{0}\right\rangle
\end{array}
\right)
\text{ ,}
\end{equation}
hence, for the mean number of $K_{S}^{0}$ and $K_{L}^{0}$ one gets:%
\begin{equation}
\label{eq:k0}
\langle K_{S}^{0} \rangle=\frac{1}{2} \langle K^{0} \rangle +\frac{1}{2} \langle \bar{K}^{0} \rangle = \langle K_{L}^{0} \rangle%
\text{ .}
\end{equation}
Thus, we have an equal number of long-lived and short-lived kaons, even if $\langle K^{0} \rangle$ and
$ \langle \bar{K}^{0} \rangle$ are different due to the non-zero baryon number of the initial state.

By combining Eqs.~(\ref{eq:kplus}) and~(\ref{eq:kminus}) and using Eq.~(\ref{eq:k0}) one finds:
\begin{equation}
\langle K^{+} \rangle + \langle K^{-} \rangle = 2 \langle K^{0}_S \rangle~,
\end{equation}
where all multiplicities are well accessible to experiments. Consequently, the exact isospin symmetry prediction for the charged-to-neutral kaon ratio in 
nucleus-nucleus collisions with $Q/B = 1/2$ reads:
\begin{equation}
\label{rk}
R_K\equiv\frac{\langle K^{+} \rangle  + \langle K^{-} \rangle}{\langle K^0 \rangle  + \langle \bar{K}^0 \rangle} = \frac{\langle K^+ \rangle  + \langle K^- \rangle}{2 \langle K_S^{0} \rangle } = 1~.
\end{equation}
This prediction serves as the reference value for testing isospin symmetry (specifically, charge symmetry) using results from the NA61/SHINE experiment~\cite{NA61SHINE:2023azp} and previous experiments; see the next section.

Including CP-symmetry breaking into the mixing modifies the mixing matrix as \cite{Thomson:2013zua}:
\begin{equation}
\left(
\begin{array}
[c]{c}%
\left\vert K^0_{S}\right\rangle \\
\left\vert K^0_{L}\right\rangle
\end{array}
\right)  =\frac{1}{\sqrt{2(1+\left\vert \varepsilon\right\vert ^{2})}}\left(
\begin{array}
[c]{cc}%
1+\varepsilon & -(1-\varepsilon)\\
1+\varepsilon & 1-\varepsilon
\end{array}
\right)  \left(
\begin{array}
[c]{c}%
\left\vert K^{0}\right\rangle \\
\left\vert \bar{K}^{0}\right\rangle
\end{array}
\right)  \text{,}%
\end{equation}
which leads to 
\begin{equation}
\left\langle K^{0}\right\rangle +\left\langle \bar{K}^{0}\right\rangle
\simeq2\left\langle K^0_{S}\right\rangle \left(  1-2~(\operatorname{Re}%
\varepsilon)~\frac{\left\langle K^{+}\right\rangle -\left\langle K^{-}\right\rangle
}{\left\langle K^{+}\right\rangle +\left\langle K^-\right\rangle }\right)
\text{ .}%
\end{equation}
In turn, the value of $R_{K}$
\begin{equation}
R_{K}=\frac{\left\langle K^{+}\right\rangle +\left\langle K^{-}\right\rangle
}{\left\langle K^{0}\right\rangle +\left\langle \bar{K}^{0}\right\rangle
}\simeq\frac{\left\langle K^{+}\right\rangle +\left\langle K^{-}\right\rangle
}{2\left\langle K^0_{S}\right\rangle }\left(  1+2~(\operatorname{Re}%
\varepsilon)~\frac{\left\langle K^{+}\right\rangle -\left\langle K^{-}%
\right\rangle }{\left\langle K^{+}\right\rangle +\left\langle K^{-}%
\right\rangle }\right)
\end{equation}
is somewhat larger than the value obtained by $(\left\langle
K^{+}\right\rangle +\left\langle K^{-}\right\rangle)~/~(2\left\langle
K^0_{S}\right\rangle)$. However, the correction is small because
$\operatorname{Re}\varepsilon\simeq1.62\cdot10^{-3} \ll 1.$ For instance, in the case of NA61/SHINE, one gets
\begin{equation}
R_{K}\simeq\frac{\left\langle K^{+}\right\rangle +\left\langle K^{-}%
\right\rangle }{2\left\langle K^0_{S}\right\rangle }\left(  1+0.001\right)~.
\end{equation} 
This modification due to weak interaction can be safely neglected.

\section{Violation of charge-symmetry invariance by known effects}
\label{sec:violation}

As discussed in Secs.~\ref{sec:isospin} and~\ref{sec:flavour}, isospin symmetry is an approximate symmetry of strong interactions.
This is because the light quark masses, $m_u \neq m_d$, are different. In addition, the symmetry is explicitly broken by weak and electromagnetic processes. 
Breaking by weak processes is negligible, as shown in the previous section. Electromagnetic effects have been discussed in 
Ref.~\cite{NA61SHINE:2023azp}. While simple estimates suggest they are small, further studies will be required.

As pointed out in Sec.~\ref{sec:charge}, the NA61/SHINE method tests the charge-symmetry invariance, which is a part of isospin symmetry. The experimental quantity used in the test is
the ratio of charge-to-neutral kaons, $R_K$.
Here, we extend the previous discussion in~\cite{NA61SHINE:2023azp} about the $R_K$ sensitivity on
effects violating the charge-symmetry invariance within strong interactions. The quantitative calculations are performed within the statistical Hadron Resonance Gas model~\cite{Vovchenko:2019pjl}. In doing so, also the fact that $Q/B<1/2$ must be taken into account. For Ar-Sc used in NA61/SHINE, $Q/B=0.459$, while for lead-lead scattering $Q/B = 0.394$. Thus, the choice $Q/B=0.4$ shall also be investigated in HRG calculations.

The section concludes with a discussion of charge-symmetry-invariance violation in the charm sector.

\subsection{Violation in the kaon sector}
\label{sec:vkaons}

\begin{figure}[htbp]
    \centering
    \begin{subfigure}[b]{0.49\textwidth}
        \includegraphics[width=\textwidth]{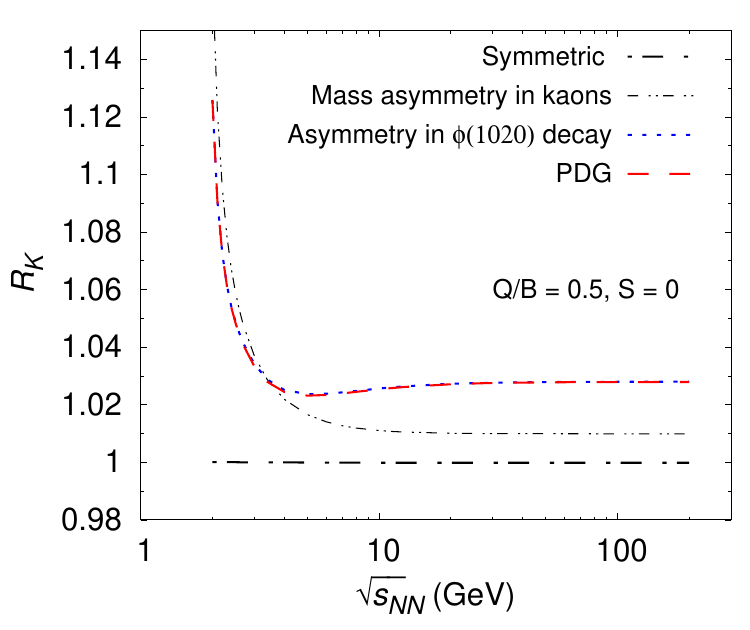}
        %\caption{xxxxx1}
        %\label{fig:subfig1}
    \end{subfigure}
    \hfill
    \begin{subfigure}[b]{0.49\textwidth}
        \includegraphics[width=\textwidth]{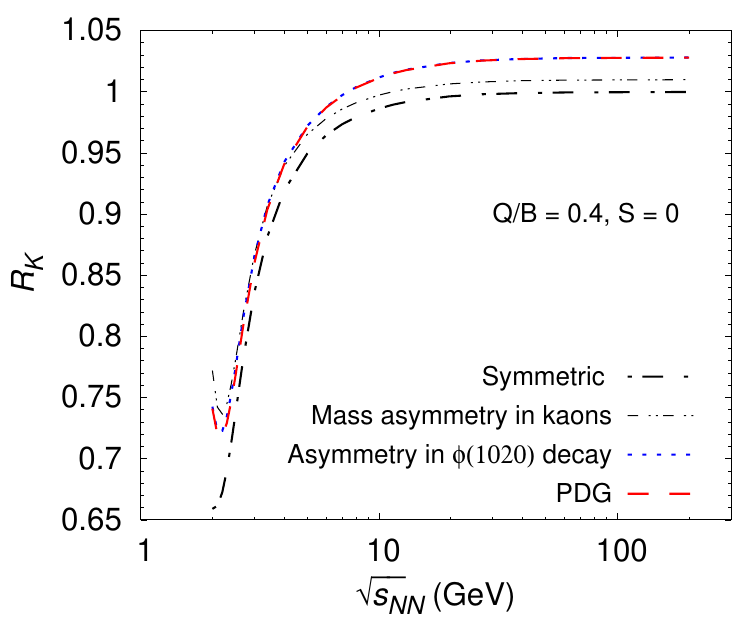}
        %\caption{xxxxx2}
        %\label{fig:subfig2}
    \end{subfigure}
    \caption{\label{fig:phi}\textit{Left:} $R_K$ as a function of $\sqrt{s_{NN}}$ calculated within HRG~\cite{Vovchenko:2019pjl}
    for $Q/B=0.5$ for different cases (from down upwards): assuming exact isospin symmetry: $R_K=1$ as expected; including PDG kaon masses; including PDG $\phi(1020)$ branching ratios; including PDG data for all particles. \textit{Right:} as in the left panel, but for $Q/B = 0.4$.}
     \label{fig:}
\end{figure}

\begin{figure}[htbp]
    \centering
    \begin{subfigure}[b]{0.49\textwidth}
        \includegraphics[width=\textwidth]{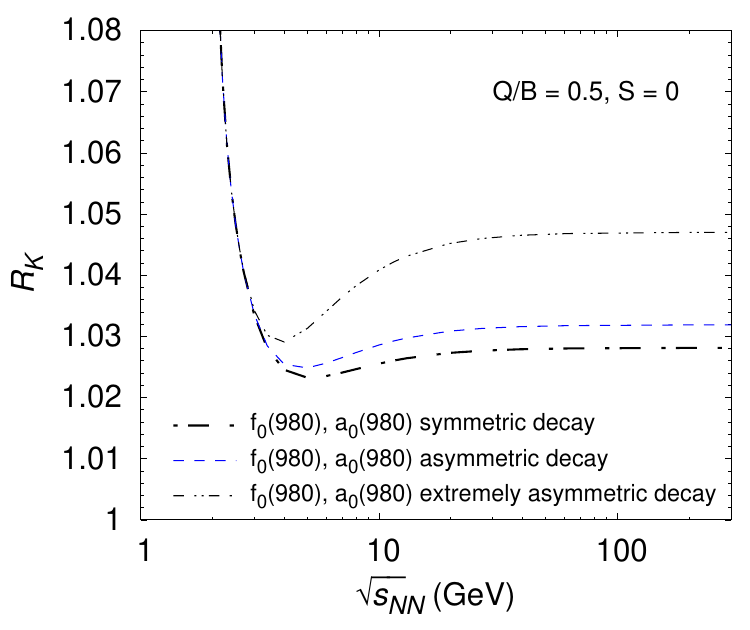}
        %\caption{xxxxx1}
        %\label{fig:subfig1}
    \end{subfigure}
    \hfill
    \begin{subfigure}[b]{0.49 \textwidth}
        \includegraphics[width=\textwidth]{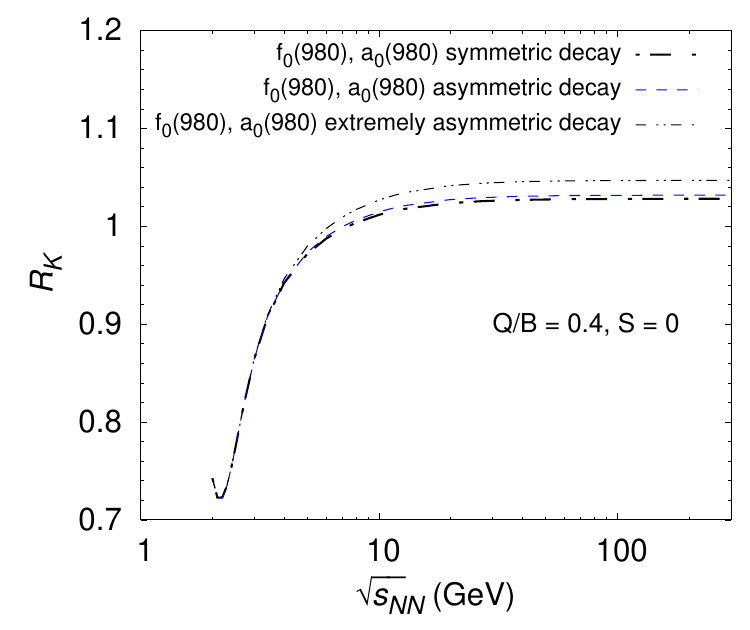}
                %\caption{xxxxx2}   
    \end{subfigure}
    \caption{\label{fig:a0f0}\textit{Left:} $R_K$ as a function of $\sqrt{s_{NN}}$ calculated within HRG~\cite{Vovchenko:2019pjl} for $Q/B=0.5$ setting different branching ratios of $a_0(980)$ and $f_0(980)$ resonances (from down upwards): $a_0(980)$ and $f_0(980)$ with isospin symmetric decays; $a_0(980)$ and $f_0(980)$ with realistic isospin breaking branching ratios; $a_0(980)$ and $f_0(980)$ with $K\bar{K}$ mode going entirely into $K^+K^-$ (unrealistic extreme choice). \textit{Right:} as in the left panel, but for $Q/B = 0.4$.}
\end{figure}

Different $u$ and $d$ quark masses violate $ud$-flavour symmetry and consequently charge-symmetry.
\iffalse
Following Sec. \ref{sec:flavour}, as the quark-mass difference with respect to the QCD mass scale  $\Lambda_{QCD}$ is small ~\cite{Kneur:2011vi,Deur:2016tte}, the ratio $R_K$ can be expanded as (assuming analytic behaviour):
\begin{equation}
    R_K = R_K^{(0)} + \frac{m_d-m_u}{\Lambda_{QCD}} R_K^{(1)} +... \text{ ,}
\end{equation}
where $R_K^{(0)}$ is the value of the ratio for the isospin symmetry ($m_d = m_u$) being equal to one for $Q/B = 0.5$.
The question is whether $R_K^{(1)} + ...$ terms can reproduce the experimental data.
\fi
We list below the considered effects and quantify their impact on $R_K$
using the HRG model~\cite{Vovchenko:2019pjl}.
In Ref.~\cite{NA61SHINE:2023azp}, the HRG results were cross-checked within the microscopic, UrQMD transport model~\mbox{\cite{Bass:1998ca,Bleicher:1999xi,Bleicher:2022kcu}}.

\vspace{0.1cm}
\begin{enumerate}[(i)]
    \vspace{0.1cm}
    \item 
    In Fig.~\ref{fig:phi}~(\textit{left}), we have verified that for $Q/B = 0.5$ the ratio $R_K =1$ is reproduced in the isospin-symmetric limit by the HRG.
    For this test, the masses of hadrons, resonances, and their decays were set to values that exactly obey isospin symmetry.
    For $Q/B = 0.4$ (the right panel), the isospin-symmetric HRG predicts an increase in $R_K$ with collision energy that reaches one above 10~GeV. More neutrons than protons (more $d$ than $u$ quarks) in the initial ensemble favours neutral kaon production.
    The effect of the initial ensemble asymmetry diminishes at high energies because of the dominance of created up and down quarks over the constituent quarks. In this limit, the theoretical predictions are very similar to the $Q/B=1/2$ case. 

    \vspace{0.1cm}
    \item 
    Smaller masses of charged kaons than neutral ones, $m_{K^+}=m_{K^-}=493.7$~MeV and $m_{K^0}=m_{\bar{K}^0}=497.6$~MeV,
    lead to an increase of $R_K$ resulting from direct kaon production by about $2\%$, see Fig.~\ref{fig:phi}. 
    This was estimated by removing resonances from the HRG particle list.
    \vspace{0.1cm}
    \item 
    Different kaon masses impact the branching ratios 
    of hadronic resonances decaying into kaons. The most relevant example is the $\phi(1020)$ meson, which decays twice as frequently to charged kaons as to neutral ones. This large difference is a consequence of the mass of $\phi(1020)$ being just above the threshold for decays into two kaons. Including the kaon production from resonance decays
    increases $R_K$ by about 4\%; see Fig.~\ref{fig:phi}.
    This figure
    also shows that the $\phi$ meson decays have the largest effect on $R_K$ among all known resonance decays.
     \vspace{0.1cm}
    \item 
      In connection to the previous point, other potentially relevant $\bar{K}K$ decays refer to the resonances $a_{0}(980)$ and $f_{0}(980)$, whose masses are close to the kaon-kaon threshold. In the PDG table for both resonances, only $\bar{K}K$ is quoted. The standard HRG model has the same branching ratios to $K^{+}K^{-}$ and $K^{0}\bar{K}^{0}$ channels. In particular, in the HRG used in this work, $a_{0}^{0}(980)$ has two decay channels: $\pi^{0}\eta$ ($85\%$) and $K\bar{K}$ ($15\%$), with $7.5\%$ for $K^{+}K^{-}$ and $7.5\%$ for $K^{0}\bar{K}^{0}$. The resonance $f_{0}(980)$ has in HRG the decay channels $\pi^{+}\pi^{-}$ ($52.7\%$), $\pi^{0}\pi^{0}$ ($26.3\%$), and $K\bar{K}$ ($21\%$), with $10.5\%$ for $K^{+}K^{-}$ and $10.5\%$ for $K^{0}\bar{K}^{0}$. Yet charge symmetry violation may be relevant for these resonances. Using Flatt\'{e}-like spectral functions as in Refs.~\mbox{\cite{Flatte:1976xv,Baru:2004xg,Giacosa:2021mbz}} and taking into account the difference of the charged and neutral kaon masses, we find for both resonances a charged-to-neutral ratio of decay widths of about $K^{+}K^{-}/K^{0}\bar{K}^{0}\simeq1.1$.
      This ratio remains smaller than $1.2$ when the parameters of the Flatt\'{e}-like functions are varied within reasonable ranges (see e.g. the compilation of couplings in Ref.~\cite{Wu:2008hx}). We thus consider the HRG approach by assigning $20\%$ more charged over neutral kaons for both resonances $a_{0}^{0}(980)$ and $f_{0}(980)$. In all cases, the ratio $R_{K}$ changes by $0.36 \%$  at large $\sqrt{s_{NN}}$ . 
      We also considered the `extreme scenario' where $a_{0}^{0}(980)$ and $f_{0}(980)$ decay only into charged kaons. It gives an upper limit of the impact of $a_{0}^{0}(980)$ and $f_{0}(980)$ decays on $R_K$. There is no significant effect
      for $\sqrt{s_{NN}}\lesssim 10$ GeV, and a slight (1.84\%) increase of $R_{K}$ at high energies, see Fig.~\ref{fig:a0f0}.

    \vspace{0.1cm}
    \item  
    Mass differences of hadrons belonging to other isospin multiplets, as well as their decays into kaons, also break the charge symmetry and affect $R_K$. 
    However, their impact is significantly smaller than the $\sim 4\%$ effect of $\phi$ decays; see Fig.~\ref{fig:phi}. 
\end{enumerate}
\vspace{0.1cm}

\subsection{Violation in the charm sector}
\label{sec:vcharm}

Similar to kaons,
exact charge-symmetry invariance predicts an equal production of charged mesons, $D^+$ and $D^-$, and neutral ones, $D^0$ and $\bar{D}^0$. However, the production of charged $D^{\pm}$ mesons in $h+A$ interactions is suppressed relative to $D^0$ and $\bar{D}^0$~\cite{Frixione:1997ma}.
A similar suppression was observed in nucleus--nucleus collisions at RHIC and at the LHC.
For example, in Au+Au collisions at $\sqrt{s_{\rm NN}} = 200\,\text{GeV}$ (10--40\% centrality), the STAR measurement finds mid-rapidity cross sections 
$d\sigma_{D^0}/dy \simeq 41\,\mu\text{b}$ and 
$d\sigma_{D^\pm}/dy \simeq 18\,\mu\text{b}$, giving a yield ratio~\cite{Vanek:2021axf,Xie:2017nal}
\begin{equation}
    R_D = \frac{\langle D^+\rangle + \langle D^- \rangle}{\langle D^0 \rangle + \langle \bar D^0 \rangle} \approx 0.5 \text{ .}
\end{equation}
At the LHC, the ALICE measurements of prompt $D^0$ and $D^+$ production in Pb--Pb collisions 
at $\sqrt{s_{\rm NN}} = 5.02\,\text{TeV}$ show the ratio $\langle D^+ \rangle /\langle D^0 \rangle \approx 0.4$--$0.5$ ~\cite{ALICE:2018lyv}, and is compatible (within uncertainties) with the same ratio in $p+p$ interactions ~\cite{ALICE:2021rxa}.

Following the same reasoning as in the kaon case, this is a manifestation of charge-symmetry-breaking. However, in the charm case, there is a simple explanation for the breaking.
The charged vector resonance $D^*(2010)^{+}$ decays into $D^+\pi^0$ and $D^0\pi^+$ with branching ratios of $30.7 \pm 0.5 \%$ and $67.7 \pm 0.5 \%$, respectively. Their ratio is close to 0.5, as expected from isospin-symmetry invariance. On the other hand, the resonance $D^*(2007)^{0}$ decays into $D^0 \pi^0 $ with branching ratio of $64.7 \pm 0.9 \%$, but the decay into its charged partner, $D^- \pi^+$, is forbidden because it is below the threshold. Exact isospin symmetry would imply a $1:2$ ratio between  $D^0 \pi^0 $ and $D^- \pi^+$ decays. The effect is qualitatively similar to $\phi(1020)$ resonance decay in the case of kaons. Quantitatively, the impact of $D^*$ decays on the charged-to-neutral $D$ ratio is much larger, since the resonance $D^*(2007)^0$ does not decay at all into $D^+$. Moreover, this resonance has a significant ($35.3 \pm 0.9 \%$) branching ratio to $D^0 \gamma$, which also increases neutral $D$ meson production w.r.t. charged $D$ mesons.
Similar considerations hold for the charge-conjugated states $D^*(2010)^{+}$ and $\bar{D}^*(2007)^{0}$. Summarising, neutral $D$ mesons are significantly favoured (by a factor of about two) over charged ones.

\begin{figure}[h!]
\centering

\resizebox{0.9\textwidth}{!}{
  \includegraphics{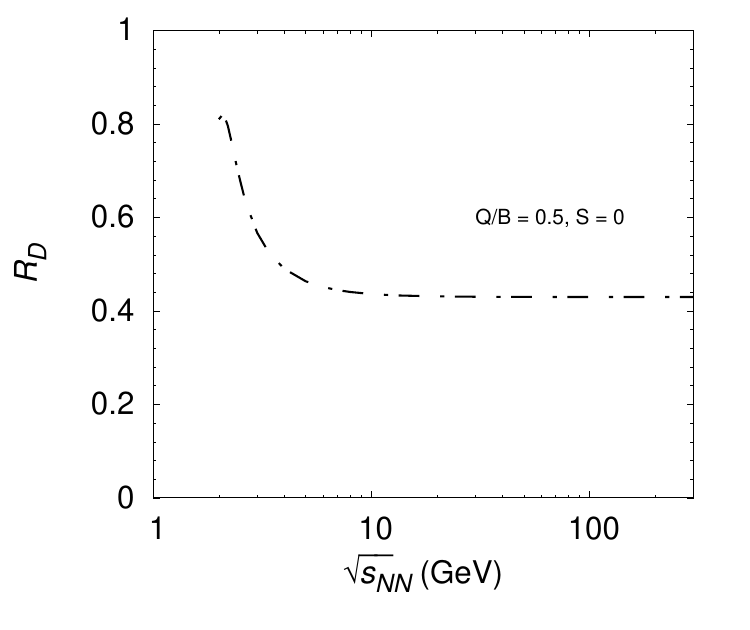}
}
\caption{Ratio $R_D$ within HRG for $Q/B=0.5$ and including resonances up to $D^*(2010)$. Note, using $Q/B=0.4$ leads to negligible numerical differences.
}
\label{fig:RD}
\end{figure}

In Fig.~\ref{fig:RD} we show the HRG result for $R_D$. 
At high energies, the model yields $ R_D\simeq 0.45$, in good agreement with the experimental value $ R_D\simeq 0.5$. 
This agreement indicates that the HRG framework provides a reliable baseline description for charmed mesons. 
The fact that the same framework fails to reproduce the charged-to-neutral kaon ratio highlights the non-trivial nature of the $R_K$ puzzle.

We conclude this section by stressing that the known effects violating charge symmetry in strong interactions cannot explain the observed violation of charge symmetry between charged and neutral kaons.
However, an even larger violation of the charged symmetry measured for
$D$ mesons can be explained by an asymmetry in $D^*$ meson decays.

\section{Closing remarks}
\label{sec:closing}

In this work, triggered by the novel experimental findings of Ref. \cite{NA61SHINE:2023azp} and discussion therein, we have investigated the ratio $R_K$ of charged-to-neutral kaon yields in nucleus-nucleus collisions at high energies, both within the general theoretical QCD framework and within the Hadron Resonance Gas model. 

In the limit of exact $ud$-flavour symmetry, the $R_K$ ratio is unity if the initial state of nuclei has an equal number of protons and neutrons ($Q/B=1/2$). Here, we have didactically clarified the origin of the result, presented a formal proof, and discussed differences to the Shmushkevich method.

The known charge symmetry violation effects included in HRG yield $R_K \approx 1.03$ at high collision energies.  
Many experimental results correspond to collisions of heavy nuclei with more neutrons than protons ($Q/B \simeq 0.4$). This imbalance reduces $R_K$ significantly
only at low collision energies, where the ratio is lower than one. We thus conclude that the known charge-symmetry-violating effects cannot explain experimental results on nucleus-nucleus collisions, for which $R_K$ is mostly within the range 1.1-1.2. 

The evidence for unexpectedly large CSV in the kaon sector is well established. 
It is, however, important that new experimental tests take place in the future.
Measurements of the charged-to-neutral kaon ratio in collisions of an equal number of protons and neutrons would reduce interpretation uncertainties. In this respect, the NA61/SHINE collaboration initiated an investigation of the $R_K$ ratio in oxygen-oxygen scattering.
In addition, measurements of $R_K$ in deuterium-deuterium interactions, the lightest nuclei with $I = 0$ and $Q/B=1/2$ would also be important. 

Another interesting system concerns pion-nucleon scattering, in particular $\pi^+  + C$ and $\pi^- + C$. Previous data on $\pi^- + C$ 
have shown that $R_K \simeq 1.2$~\cite{NA61SHINE:2022tiz}, in line with nucleus-nucleus scattering. Charge symmetry implies that the average $R_K$ obtained by combining both measurements should be close to unity. On the other hand, the coalescence approach for hadronic production~\cite{Giacosa:2025ynn} shows that both systems should lead to the same enhanced $R_K$ with $R_K^{\pi^+  C} =R_K^{\pi^- C} \simeq 1.2 >1$.

On the theoretical side, different directions can be taken. 
Concerning the HRG approach, one may argue that isospin-breaking contributions have not yet been included. This scenario seems, however, improbable: the higher the mass of a resonance, the smaller its multiplicity. Moreover, such resonances should exhibit a large, unexplained isospin-symmetry-breaking decay to account for the experimentally measured $R_K$.

One may also consider models that parametrise isospin-breaking processes and fit their parameters to the data. 
Along this line, the UrQMD approach has been modified by including violation of isospin-symmetry~\cite{Reichert:2025znn}: the string breaking is parametrized as three times more probable to produce $\bar{u}u$ pair than a $\bar{d}d$ one; in this way, the ratio $R_K$ as well as BESIII data on kaon productions in $\e^+ e^-$ scattering can be correctly described. 
Similarly, the quark coalescence model of Ref.~\cite{Giacosa:2025ynn} has been used to fit to $R_K$ data, implying an excess of $u/d$ quarks, and to make predictions such as the $p/n \simeq 1.2$ and $\Sigma^+/ \Sigma^- \simeq 1.4$ ratios.

It is expected that, in the near future, both experimental and theoretical advances will help clarify the origin of the isospin kaon anomaly.

\vspace{0.5cm}
\textbf{Acknowledgments}

The authors thank Wojciech Broniowski, Katarzyna Grebieszkow, Stanislaw Mrowczynski, Owe Philippsen, Adam Para, Rob Pisarski, Krishna Rajagopal, Jan Steinheimer, Leonardo Tinti, Ludwik Turko, and Volodymyr Vovchenko for fruitful discussions and comments.  A special thank you goes to  Herbert Stroebele for suggestions, corrections, and remarks.
This work is partially supported by the Polish National Science Centre (NCN) grants 2018/30/A/ST2/00226 and 2019/33/B/ST2/00613 and the Alexander von Humboldt Foundation. 

\newpage
\bibliographystyle{unsrt} % any style is allowed; include your .bib
\bibliography{references}

\end{document}